\documentclass[%
prx,
 amsmath,amssymb,
 aps,
twocolumn,
longbibliography,
showkeys
]{revtex4-1}

\usepackage{graphicx}
\usepackage{epstopdf}
\usepackage{dcolumn}
\usepackage{bm}
\usepackage{color}
\usepackage[mathlines]{lineno}
\usepackage[breaklinks=true,colorlinks=true,linkcolor=blue,urlcolor=blue,citecolor=blue]{hyperref}
\usepackage{ragged2e}
\usepackage{enumitem}
\usepackage{bibunits}
\usepackage{amsthm}
\usepackage{diagbox}

\begin{document}


\title[]{Mixed strategy approach destabilizes cooperation in finite populations with clustering coefficient}

\author{Zehua Si$^1$}
\author{Zhixue He$^{2,1}$}
\author{Chen Shen$^3$}
\email{steven\_shen91@hotmail.com}
\author{Jun Tanimoto$^{1,3}$}
\email{tanimoto@cm.kyushu-u.ac.jp}

\affiliation{
\vspace{2mm}
\mbox{1. Interdisciplinary Graduate School of Engineering Sciences, Kyushu University, Fukuoka, 816-8580, Japan}
\mbox{2. School of Statistics and Mathematics, Yunnan University of Finance and Economics, Kunming, 650221, China}
\mbox{3. Faculty of Engineering Sciences, Kyushu University, Kasuga-koen, Kasuga-shi, Fukuoka 816-8580, Japan}
}

\date{\today}

\begin{abstract}

Evolutionary game theory, encompassing discrete, continuous, and mixed strategies, is pivotal for understanding cooperation dynamics. Discrete strategies involve deterministic actions with a fixed probability of one, whereas continuous strategies employ intermediate probabilities to convey the extent of cooperation and emphasize expected payoffs. Mixed strategies, though akin to continuous ones, calculate immediate payoffs based on the action chosen at a given moment within intermediate probabilities. Although previous research has highlighted the distinct impacts of these strategic approaches on fostering cooperation, the reasons behind the differing levels of cooperation among these approaches have remained somewhat unclear. This study explores how these strategic approaches influence cooperation in the context of the prisoner's dilemma game, particularly in networked populations with varying clustering coefficients. Our research goes beyond existing studies by revealing that the differences in cooperation levels between these strategic approaches are not confined to finite populations; they also depend on the clustering coefficients of these populations. In populations with nonzero clustering coefficients, we observed varying degrees of stable cooperation for each strategic approach across multiple simulations, with mixed strategies showing the most variability, followed by continuous and discrete strategies. However, this variability in cooperation evolution decreased in populations with a clustering coefficient of zero, narrowing the differences in cooperation levels among the strategies. These findings suggest that in more realistic settings, the robustness of cooperation systems may be compromised, as the evolution of cooperation through mixed and continuous strategies introduces a degree of unpredictability.

\end{abstract}


\maketitle

\section{Introduction}       
Cooperation among unrelated individuals is a widespread phenomenon in both natural and human societies. While cooperation typically incurs costs to benefit others, it paradoxically persists despite the evolutionary advantage of free riding, which avoids the costs associated with helping others. This apparent contradiction challenges the principle of `survival of the fittest'~\cite{darwin2023origin}, raising critical questions about how cooperation evolves and sustains itself. Addressing this question has been the focus of multidisciplinary efforts over the past decades. Evolutionary game theory, initially formalized by Maynard Smith and George Price~\cite{smith1973logic,smith_1982}, offers a robust mathematical framework to analyze this issue. Within this framework, various mechanisms of reciprocity, including direct~\cite{van2012direct,hilbe2018partners}, indirect~\cite{nowak1998dynamics,leimar2001evolution,nowak2005evolution}, and network reciprocity~\cite{wang2013interdependent,nowak1992evolutionary}, as well as group and kin selection~\cite{lehmann2007group,van2009group}, have been proposed to explain the emergence and maintenance of cooperation. The common principle underlying these mechanisms is positive associativity, which necessitates the ability of individuals to recognize cooperators, enabling cooperation with fellow cooperators and defection against defectors. 

Traditionally, evolutionary game theory has emphasized a discrete strategy approach, wherein individuals are faced with a binary choice: either to cooperate (benefiting others at a personal cost) or to defect (free-riding on others' efforts). However, observations from biological species suggest that individual strategies can be more nuanced and continuous. A notable example is the Harris's sparrow, which signals its social dominance through the varying color intensity of its head and throat feathers\cite{rohwer1977status}. Acknowledging this, Maynard Smith proposed the continuous strategy approach~\cite{smith_1982}. This approach allows for a spectrum of choices between cooperation and defection, represented by a probability continuum ranging from 0 to 1. In this context, an individual's strategy is expressed as a probability, reflecting the degree of cooperation, and their fitness is determined by the expected payoff. In parallel, the mixed strategy approach, akin to the continuous strategy but distinct, involves making choices with a certain probability~\cite{crawford1985learning,haccou1995optimal,szabo2016role}, where fitness is determined by the immediate payoff of the action chosen at that moment, rather than an expected payoff.

Compared with discrete strategic approach, continuous and mixed strategic approaches have received less attention in the past. While some studies have independently investigated the impact of continuous and mixed strategic approaches on cooperation~\cite{killingback2002continuous,wahl1999continuous,zhao2015evolution,wang2020learning,han2022hybrid,zhang2021role,wang2019roles,kokubo2015spatial}, there has been relatively little exploration of the differences or similarities in their effects on cooperation dynamics. Examining the disparities between these strategic approaches, existing research indicates that cooperation levels tend to be higher under continuous strategic approaches, especially in finite and well-mixed populations when the payoff function of the continuous strategy approach is the linear interpolation of the payoff matrix of the discrete strategy. This tendency is particularly amplified when network structures are taken into account~~\cite{zhong2012equilibrium}. Furthermore, Kokubo et al. conducted a comprehensive comparison of equilibrium states across these three strategic approaches within structured populations. Their findings revealed that boundary games, such as chicken and stag hunt, exhibit distinct mechanisms that lead to substantial differences in equilibrium outcomes when comparing continuous or mixed strategy games to discrete strategy games\cite{kokubo2015spatial}. This divergence in outcomes has also been observed in coevolutionary models, where the introduction of mixed strategies in a spatial game context was found to significantly enhance cooperation when contrasted with conventional discrete or continuous strategies~\cite{tanimoto2017coevolution}. Additionally, Wang et al. delved into the particle swarm optimization strategy (PSO) update rules based on continuous and mixed strategy systems, respectively, within scale-free networks. Their findings indicate that the PSO strategy update rules can promote cooperation in the continuous strategy system, but may lead to a significant decline in cooperation within the mixed strategy system\cite{wang2019roles} 

Despite these findings, one question remains: Why does cooperation vary among these three strategic approaches given that such approaches do not alter the game's equilibrium? To address this, we examined the influence of these strategies on cooperation within spatial prisoner's dilemma games, both in well-mixed populations (depicted through complete graphs) and in networked populations.  Our findings initially affirm that differences in these strategic approaches manifest primarily in finite-size populations and diminish as population size increases. Across various simulations, we observe varied levels of stable cooperation for each strategic approach, with the mixed strategy approach exhibiting the greatest uncertainty, followed by continuous and discrete strategies, respectively. This pattern persists in both well-mixed and networked populations. However, an exception occurs in networked populations with a clustering coefficient of zero, where we noted a convergence in cooperation levels across various independent simulations for each strategic approach, even within finite-size populations. Such findings suggest that the differences in cooperation levels can be attributed to the stochastic evolution of cooperation. While the mixed and continuous strategies exhibit greater variability in cooperation's evolution, the introduction of clustering coefficients plays a pivotal role in mitigating this variability, alongside the influence of an increasing population size.

\section{Model}
The core of our model approach consists of four components: (i) discrete, continuous, and mixed strategies setups; (ii) population structure; (iii) strategy update; and (iv) simulation settings. Next, we provide a brief description of each of these four components.
\subparagraph{(i) Discrete, continuous, and mixed strategies setups}
We consider spatial prisoner’s dilemma ($SPD$) game as a paradigm. Following that, we will proceed with delivering a segmented description of these three strategies:

 \begin{figure*}[!t]
    \centering
\includegraphics[width=0.86\linewidth]{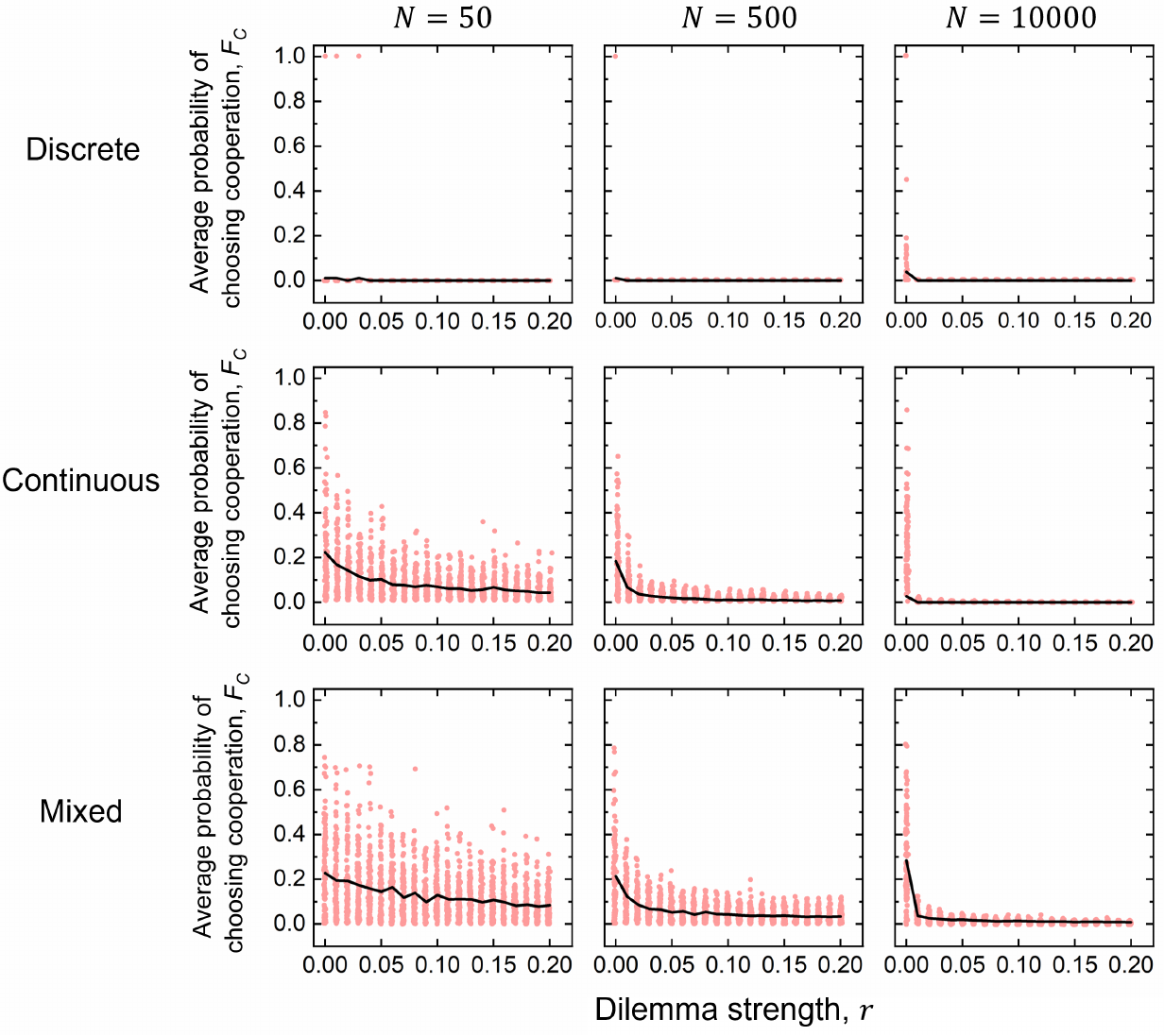}
      \caption{The probability of choosing cooperation ($F_c$) in well-mixed populations, obtained from 100 independent simulations, is depicted across three strategy settings for populations of sizes $N=50$ (left panel), $N=500$ (middle panel), and $N=10,000$ (right panel), as a function of the dilemma strength ($r$). Results are presented for discrete, continuous, and mixed strategic approaches from top to bottom, respectively. Stable cooperation levels from each simulation are represented by pink dots, while the black solid line indicates their average. For smaller population sizes, all three strategic approaches show varying degrees of uncertainty regarding stable cooperation levels, with the mixed strategy scenario displaying the highest uncertainty, followed by continuous and discrete strategies. Increasing population size mitigates this uncertainty, as evidenced by the convergence of cooperation levels across each independent simulation for every strategic approach. }
    \label{fig1}
\end{figure*}
\paragraph{Discrete strategy}
In discrete strategy, players choose either pure cooperation ($C$) or pure defection ($D$), with probability of 1. Mutual cooperation results in a reward $R$ for both parties, while mutual defection incurs a punishment $P$. If one player chooses cooperation while the other opts for defection, the former gains the sucker's payoff $S$, while the latter obtains the temptation payoff $T$. According to the universal dilemma strength proposed in Refs.~\cite{wang2015universal,tanimoto2021sociophysics} , we defined the chicken-type dilemma as $D_g = T-R$ and the stag hunt-type as $D_r = P-S$. For the sake of simplicity and without loss of generality, we assumed $P = 0$ and $R = 1$, then the payoff matrix can be expressed in the form of Eq.~\ref{eq1}:

\begin{equation}
{
\left[ \begin{array}{cc}
R & S \\
T & P 
\end{array} 
\right ]}={
\left[ \begin{array}{cc}
1 & -D_r \\
1+D_g & 0 
\end{array} 
\right ]}.
\label{eq1}
\end{equation}
Where the parameters satisfy $T > R > P > S$ and $2R > T+S$. We set $D_g=D_r=r\in[0,1]$ to simplify the parameters, then we can characterize the dilemma strength of $SPD$ game using a single parameter, $r$.

\paragraph{Continuous strategy}
Unlike discrete strategy approach where players can only choose either $C$ or $D$ with a probability of 1, continuous strategy allow for the selection of $C$ with an intermediate probability $s\in[0,1]$ (with $1-s$ for choosing $D$). Since the action set in this scenario is continuous rather than binary,
the payoff of player $i$ as shown in Eq.~\ref{eq2} can be calculated by the expected payoff according to the payoff matrix provided in~\ref{eq1}.

\begin{equation}
\begin{split}
\pi_{ij}=-D_r\cdot s_i+(1+D_g)\cdot s_j
&+(-D_g+D_r)\cdot s_i\cdot s_j\\
&=-r\cdot s_i+(1+r)\cdot s_j
\end{split}
\label{eq2}
\end{equation}

\paragraph{Mixed strategy}Similar to continuous strategy, mixed strategy still allow players to have an intermediate probability range from 0 to 1 as their strategy $s$. However, unlike continuous strategy where $s$ represents both the player's strategy and action, mixed strategy require players to make a definite choice between $C$ and $D$ based on their strategy. Consequently, the players' set of actions, like discrete strategy, is binary rather than continuous, and thus the calculation of their payoffs is the same as with discrete strategy (see Eq.~\ref{eq1}).


\subparagraph{(ii) Population structures}
We hypothesize two types of populations: well-mixed and networks. In the case of the former, we assumes there are $N$ players in the population, each of whom can interact with the other $N-1$ players. Owing to the characteristics of well-mixed populations, we employ a fully connected network as a representation. This is because, in a fully connected network, every node is connected to all other nodes, mirroring the interactive nature of well-mixed population. In the case of the latter, players can only interact with their nearest neighbors. Our study starts with a regular two dimensional (2D) lattice network with 8-nearest neighbors (up, down, left, right, upper left, upper right, lower left, and lower right), which means the network has an average degree $\left\langle k\right\rangle$ of 8. The network (population) size is $N$. To ensure that all nodes in the network have an equal number of interacting neighbors, we consider the network to have periodic boundaries.

\begin{figure*}[!t]
    \centering
\includegraphics[width=0.86\linewidth]{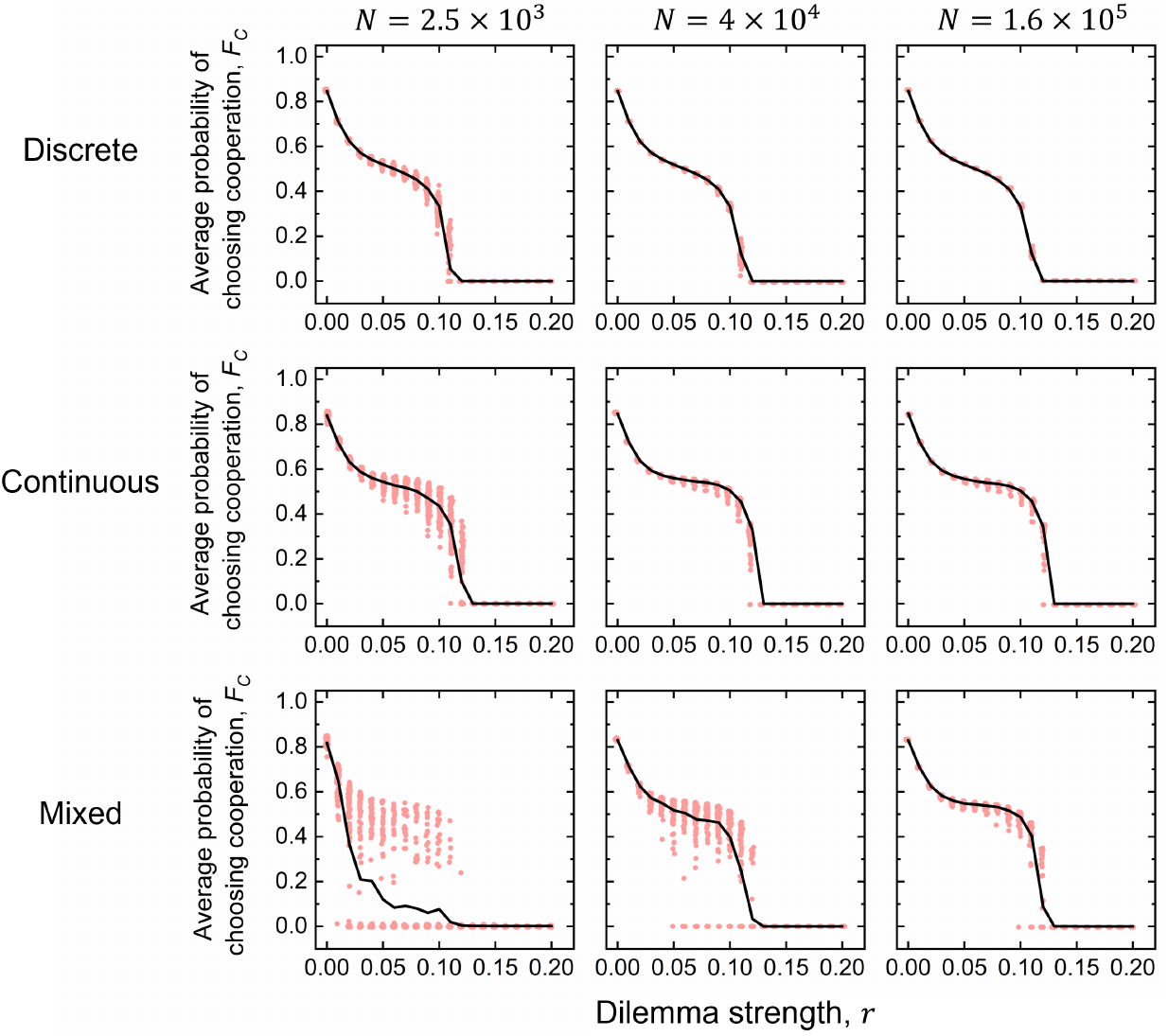}
      \caption{The probability of choosing cooperation in regular lattice, $F_c$, obtained from 100 simulations, is shown across three strategy settings in populations of size $N=2.5 \times 10^{3}$ (left panel), $N=4 \times 10^{4}$ (middle panel), and $N=1.6 \times 10^{5}$ (right panel), plotted as a function of the dilemma strength, $r$. From top to bottom, results are obtained from discrete, continuous, and mixed strategic approach, respectively. Pink dots represent stable cooperation levels from each independent  simulations, while the black solid line indicates their average. Notably, similar to results in well-mixed populations, uncertainties in stable cooperation levels are observed for each strategic approach in small-sized regular lattices, with these differences tending to converge as network size increases.}
    \label{fig2}
\end{figure*}

To probe into scenarios with differing global clustering coefficient $C_i$, we conducted simulations on networks reconstructed according to the following methodology: we randomly rewired the edges of a regular 2D lattice network, initially characterized by a significantly high $C_i$, with a rewiring probability $p$. During the rewiring process, all nodes still maintain an average degree of 8. As $p$ increases, we obtain a small-world network retaining a high $C_i$. Further increment of $p$ to 1 results in a random network characterized by a lower $C_i$. Hence, an increase in $p$ will generate networks with progressively smaller $C_i$. However, it is important to note that the above conclusion is only valid when $\left\langle k\right\rangle> 4$, for $\left\langle k\right\rangle\leq 4$, such random rewiring does not alter the network's $C_i$, with its $C_i$ always remaining equal to 0. Therefore, in addition to the scenario with $\left\langle k\right\rangle=8$, we also followed the above steps for network reconstruction in scenarios with $\left\langle k\right\rangle=4$ and 12 to compare the simulation results.

\begin{figure*}[!t]
    \centering
    \includegraphics[width=0.96\linewidth]{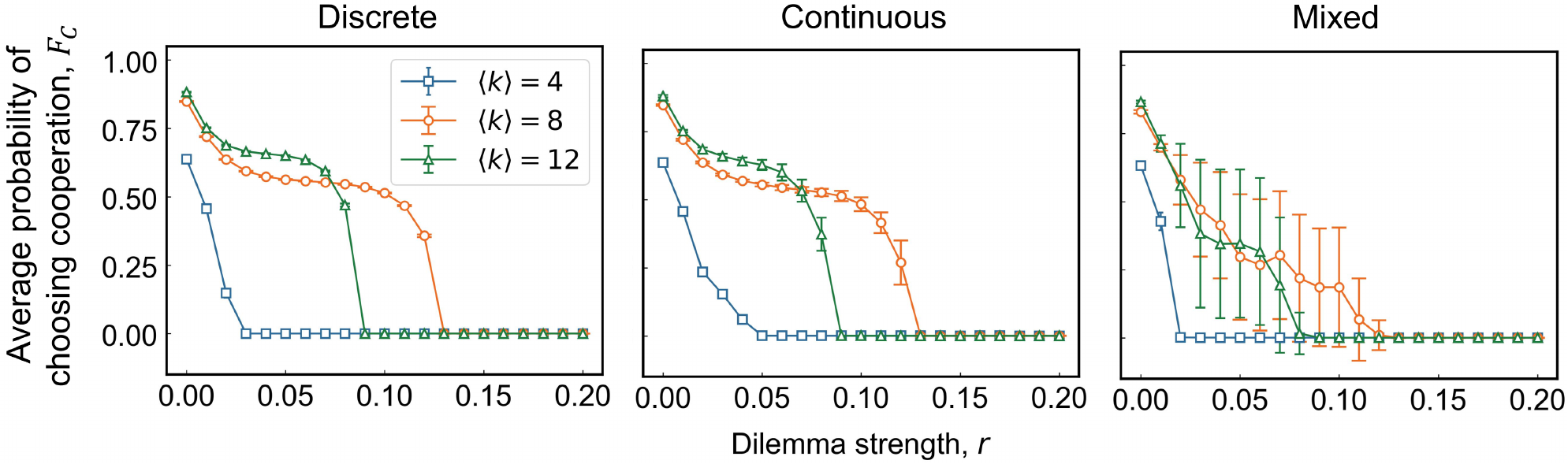}
      \caption{The figure displays the average probability of selecting cooperation, derived from 100 independent simulations, for three strategic approaches: discrete (left), continuous (middle), and mixed (right), as a function of the dilemma strength ($r$), on a regular lattice. Simulations were conducted on three types of lattices characterized by their average degrees—4 (blue) with a clustering coefficient of 0, 8 (yellow) with a non-zero clustering coefficient, and 12 (green) with a higher clustering coefficient than that of degree 8. The vertical lines represent error bars, indicating the standard deviation from these 100 simulations, to show variability in the outcomes. The employed lattice is $N=10^{4}$. It can be observed that while the mixed strategic approach exhibits considerable uncertainty in stable cooperation levels (as evidenced by the standard deviation) compared to the continuous and discrete strategic approaches for regular lattices with degrees of 8 or 12, such uncertainty disappears for regular lattices with a degree of 4, even when restricted to small lattice sizes.}
      \label{fig3}
\end{figure*}

\subparagraph{(iii) Strategy update}
We employ the pair-wise Fermi (PW-Fermi) rule as the strategy updating mechanism for players, as it allows for the characterization of human decision-making with incidental irrationality by considering the impact of noise~\cite{szolnoki2017alliance,yamauchi2010controls,shi2022coupling}. It assumes that player $i$ adopts strategy $s_i$ and engages in a game with their $k$ nearest neighbors, resulting in cumulative payoff $\pi_i$. Subsequently, player $i$ randomly selects player $j$ from their neighbors as their opponent. Player $j$ repeats the aforementioned game steps and obtains their own cumulative payoff $\pi_j$. Next, player $i$ decide whether to imitate the strategy of their opponent $j$ based on the following probability $Q$:
\begin{equation}
    Q_{s_i\gets{s_j}}=\frac{1}{1+exp(\frac{\pi_i-\pi_j}{\kappa})}
\end{equation}
where $\kappa$ represents the amplitude of stochastic uncertainties (noise) which allows players to make irrational choices. As $\kappa\rightarrow0$, which corresponds to the case of strong selection, players strictly update their strategies based on the relative payoffs. However, when $\kappa\rightarrow +\infty$, i.e., the case of weak selection, the decision of whether to update the strategy becomes highly random, akin to flipping a coin. Here we fixed $\kappa = 0.1$ to study the scenario of strong selection, as in most instances, human decision-making is predominantly rational.

\subparagraph{(iv) Simulation settings}
Monte Carlo simulations (MCS) were conducted on different networks (populations) ranging from $N=500$ to $1.6\times10^5$ to study the impact of network (population) size on the results. Each MCS gives a chance for every player to enforce its strategy onto one of the neighbors once on average. Results of simulations presented below were obtained by averaging out the final $2\times10^3$ steps after the $10^4$ time steps, ensuring the systems reached a stationary state after sufficiently long relaxation times. The final results were averaged over up to 100 independent realizations for each set of parameters to guarantee the sufficient accuracy.
 
\begin{figure*}[!t]
    \centering
    \includegraphics[width=0.96\linewidth]{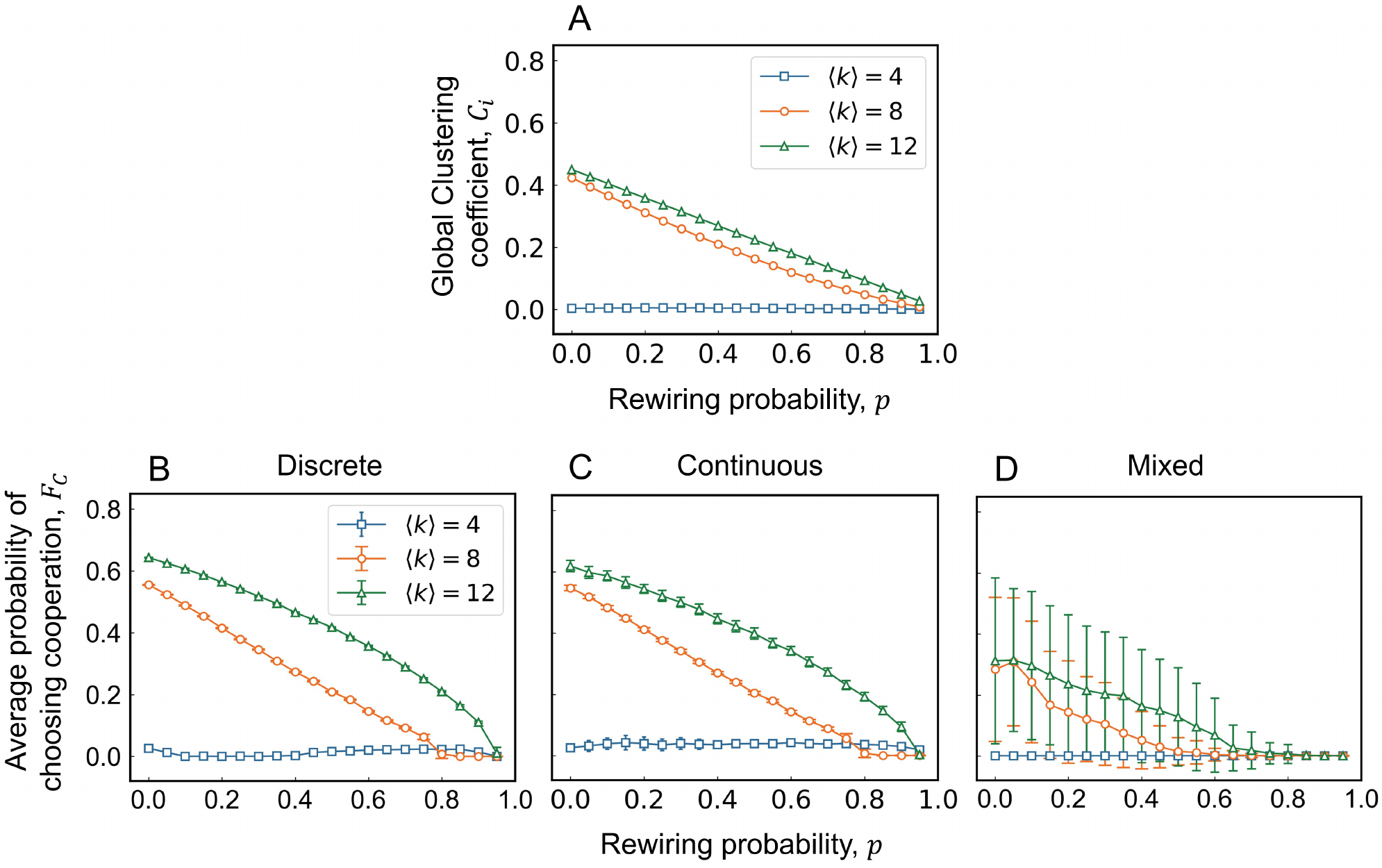}
      \caption{The top panel displays clustering coefficients as functions of the rewiring probability $p$. The bottom panel shows the average probability of choosing cooperation, derived from 100 realizations, as a function of the rewiring probability $p$ for three strategic approaches: discrete (left), continuous (middle), and mixed (right), in networked populations. Each panel presents results for degree 4 (blue), 8 (yellow), and 12 (green). Vertical lines represent error bars, indicating the standard deviation from these 100 simulations, demonstrating variability in outcomes. Results were obtained with a dilemma strength of $r = 0.05$ and network size $N=10^{4}$. Large rewiring probabilities correspond to low clustering coefficients. It can be observed that considerable uncertainty in stable cooperation levels in the mixed strategic approach diminishes for populations with clustering coefficients of 0, even though the networks remain homogeneous with degrees of 8 or 12.}
      \label{fig4}
\end{figure*}

\section{Results}
Figure \ref{fig1} presents the stable cooperation levels for each 100 independent simulation, shown as the average probability of cooperation in the stable state against the dilemma strength $r$ in well-mixed populations. The results are displayed in a top-to-bottom sequence for the discrete, continuous, and mixed strategic approaches, respectively. Horizontally, the figures are organized by network sizes: $N=50$, $N=500$, and $N=10,000$. Notably, the discrete strategy is defined by deterministic mechanisms in action selection and payoff calculation, whereas the mixed strategy introduces stochasticity in action selection, affecting the randomness of payoff calculations. The continuous strategy, akin to the mixed strategy, allows for an intermediate level of cooperation with a larger strategy space than the discrete strategy, yet it maintains determinism in action choices and the calculation of expected payoffs. 
This distinction in strategic methodologies leads to varied patterns in the evolution of cooperation under each approach, as reflected by the varied levels of cooperation in each set of independent simulations for a given value of $r$. 
In these scenarios, the variation or uncertainty of stable cooperation levels in the mixed strategic approach surpasses that in the continuous approach, which, in turn, exceeds that observed in the discrete approach. However, as the population size increases, specifically in the case of $N=10,000$ shown in the right panel of Figure \ref{fig1}, the variation in stable cooperation levels across the 100 simulations diminishes significantly. In this context, the average cooperation level across the three strategic approaches becomes nearly indistinguishable, with the exception occurring at $r=0$. At this point, the game deviates from the traditional prisoner's dilemma, as both cooperation and defection yield the same payoff, resulting in random evolutionary outcomes.

In networked populations, our observations mirrored those observed in well-mixed populations. We observed distinct evolutionary outcomes for each strategic approach in multiple independent simulations, primarily noticeable in small population sizes, as illustrated in the left and middle panels of Figure \ref{fig2}. As population sizes increased, as demonstrated in the right panel of Figure \ref{fig2}, these differences in outcomes became less pronounced. The only exception distinguishing it from well-mixed populations was that, in small population sizes, the mixed strategic approach consistently resulted in the lowest average cooperation across 100 simulations, while cooperation levels in the discrete and continuous approaches could not be reliably distinguished. 

It is important to note that the results discussed previously were derived from simulations on a square lattice with an average degree of 8. To further explore the impact of network degree on the variation in stable cooperation levels across independent simulations for three distinct strategic approaches, we conducted simulations on a small-sized regular lattice ($N=10^{4}$). Figure \ref{fig3} illustrates the average probability of choosing cooperation over 100 independent simulations, along with the corresponding standard deviations (indicated by vertical lines), for discrete (left), continuous (middle), and mixed (right) strategic approaches as functions of the dilemma strength $r$. Each panel presents outcomes for lattices with degrees of 4, 8, and 12, respectively. It was observed that lattices with a degree of 12 exhibit a similar level of uncertainty in stable cooperation—indicated by large standard deviations—as those with a degree of 8. However, this uncertainty was significantly reduced for lattices with a degree of 4, even though all results were derived from regular lattices with a small size. These findings suggest that clustering coefficients could be a key factor driving the observed variations in cooperation across independent simulations, as these networks vary in their clustering coefficients: a lattice with a degree of 8 has a non-zero clustering coefficient; a lattice with a degree of 4 has a clustering coefficient of 0; and a lattice with a degree of 12 has a higher clustering coefficient than that of a degree of 8. 

\begin{figure}[!t]
    \centering
\includegraphics[width=0.96\linewidth]{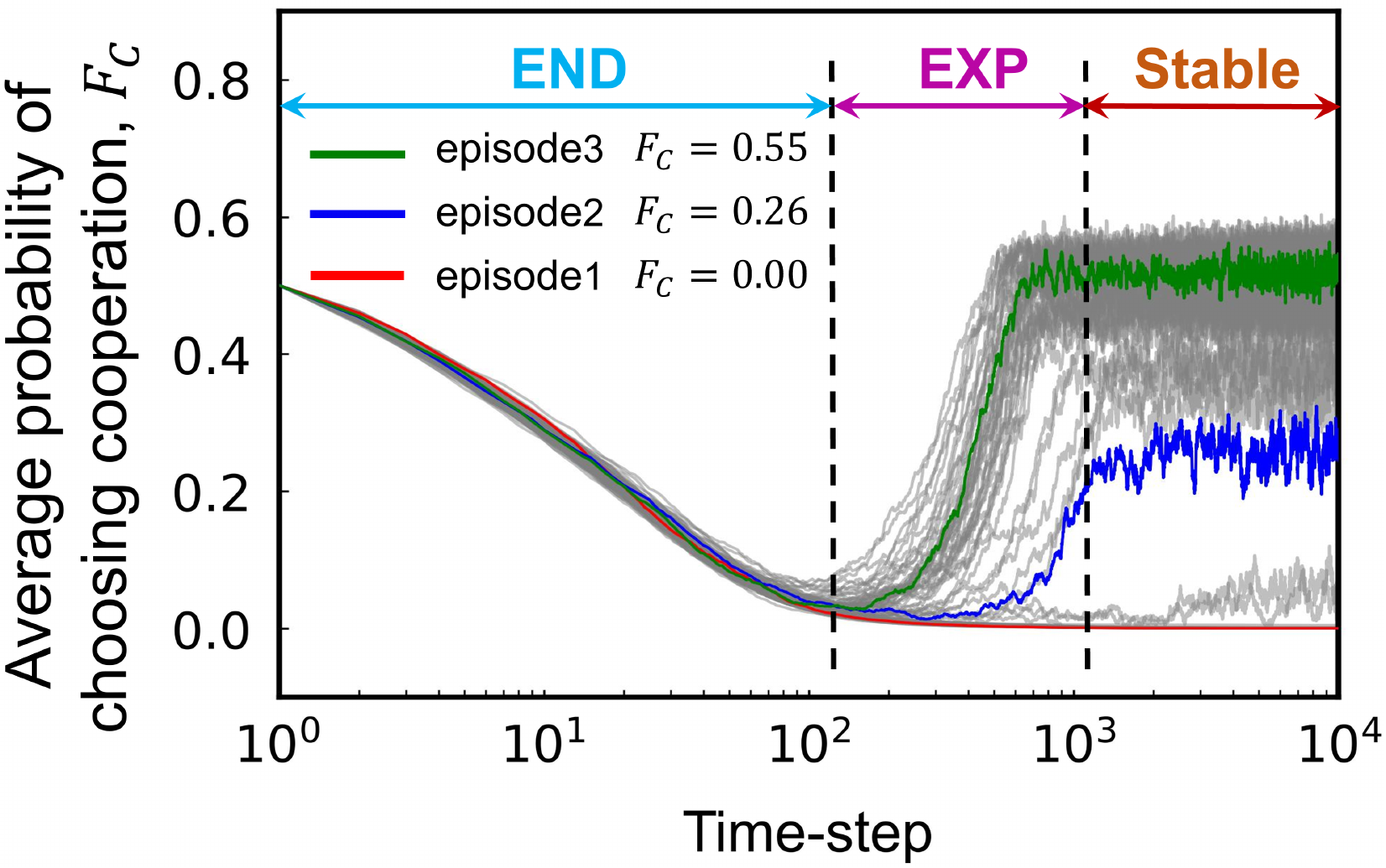}
      \caption{The average probability of selecting cooperation ($F_C$)  in the population over time under a mixed strategic approach, for each independent simulation, with the dilemma strength fixed at $r=0.05$. Shown are the time evolution diagrams of 100 independent simulations under the same parameter conditions, with three evolutionary trajectories highlighted in different colors, each ultimately stabilizing to one of the three representative evolutionary stable states. Initially, $F_C$ starts at 0.5 and exhibits a consistent downward trend during the END period in all simulations. However, during the EXP period, there is notable variability in the rate of increase of $F_C$ among different simulations, resulting in diverse stabilization levels of $F_C$ at their respective equilibrium states. The employed network size is $N=10^{4}$.}
      \label{fig5}
\end{figure}

\begin{figure*}[!t]
    \centering
    \includegraphics[width=0.91\linewidth]{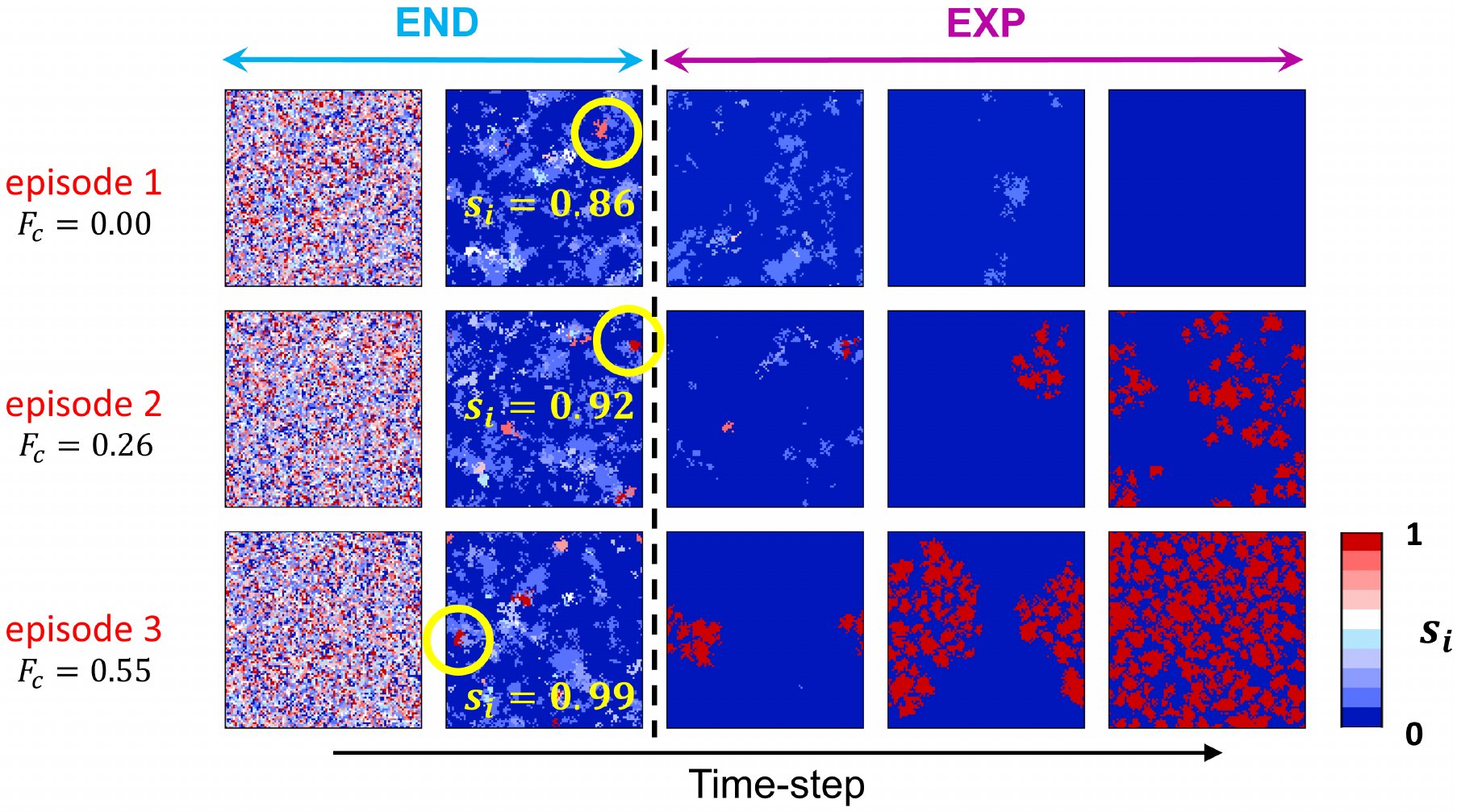}
      \caption{Evolutionary snapshots captured at different time steps for three representative episodes selected from 100 independent simulations (corresponding to those in Fig.~\ref{fig2}), with a fixed dilemma strength of 0.05. Initially, the probability of each player choosing cooperation is entirely random (first column), while the average probability of cooperation among surviving cooperative clusters varies by the end of the END period (second column). Higher probabilities of cooperation among surviving cooperative clusters at the END period correlate with greater ability to expand their territories during the EXP period (third and fourth columns), ultimately resulting in higher levels of cooperation in the stable state (last column). The employed network size is $N=10^{4}$.}
      \label{fig6}
\end{figure*}

To further substantiate the influence of clustering coefficients on cooperation, we maintained a constant average network degree while generating a spectrum of networks with varying clustering coefficients. This was achieved by rewiring the regular lattice to produce both small-world and random regular graphs, employing rewiring probabilities ranging from 1\% (yielding a small-world network) to 99\% (resulting in a random network). As illustrated in Figure \ref{fig4}A, an increase in the rewiring probability corresponded to a decrease in the clustering coefficient.

Subsequent analyses, as presented in Figures \ref{fig4}B-D, examined the average probability of choosing cooperation against the rewiring probability for discrete, continuous, and mixed strategic approaches. By keeping the network degree constant and focusing on small-sized networks, we observed minimal variability in the cooperation levels across multiple independent simulations for both discrete and continuous strategies, across the entire span of rewiring probabilities. In contrast, the mixed strategic approach demonstrated significant variations in cooperation levels for networks with degrees of 8 and 12. However, this variability diminished with a decrease in clustering coefficients (an increase in rewiring probability). For networks with a degree of 4, which consistently have a clustering coefficient of zero, the standard deviation of cooperation levels across multiple simulations was negligible, regardless of the rewiring probability. 

The results from both well-mixed populations, as indicated in Figure \ref{fig1}, and networked populations, as shown in Figures \ref{fig3} and \ref{fig4}, suggest that the considerable uncertainty in cooperation levels across multiple simulations under each strategic approach is influenced by the clustering coefficients of the populations, in addition to population size. Populations with a clustering coefficient of 0 can also mitigate the uncertainty in the evolution of cooperation under these strategic approaches. Regarding to average cooperation level among these three strategic approaches, in small-sized populations of clustering coefficient of 0, we observe that differences still persist among these three approaches, with cooperation extinction thresholds of $r=0.03$, $r=0.05$, and $r=0.02$ for the discrete, continuous, and mixed strategic approaches, respectively ( see the scenarios for average degree of 4 in figure \ref{fig3}). However, we note that  average cooperation level significantly converges across these three strategic approaches. For instances,  in scenarios that average degree of 4 in figure \ref{fig3},  cooperation levels at $r=0$ and $0.01$ are almost similar, while cooperation level in the range of [0.02, 0.09] of $r$ exhibit significant difference especially compared with mixed strategic with other strategic approaches for scenarios of networked populations with average degree of 8 or 12; in the limit of $p \to 1$ (coefficient coefficients of the population approaches to 0), cooperation level converges to 0 even for the networked populations with average degree of 8 or 12, while cooperation levels exhibit significant different for these strategic approaches if clustering coefficients of the populations with average degree of either 8 or 12 is not 0 (see bottom panel of figure~\ref{fig4}).

To further investigate the significant variations in stable cooperation levels observed in each independent simulation resulting from the mixed strategic approach, we conducted an analysis of cooperation dynamics across both temporal and spatial dimensions. We examined the temporal evolution of cooperation through time-series data and evolutionary snapshots. At the heart of the challenge in understanding cooperation lies the questions of how cooperation can persist amidst a prevalence of defectors and how it can maintain stability once established. To provide a comprehensive insight into these issues, we divided the cooperation dynamics into three distinct periods, following the framework proposed by the reference~\cite{wang2013insight,tanimoto2015fundamentals}. These periods include: the enduring period (END), characterized by cooperators forming compact clusters to resist defection invasion; the expanding period (EXP), where cooperative clusters expand their influence; and the stability period, during which the overall cooperation level stabilizes around a certain value.

Figure \ref{fig5} illustrates the average probability of choosing cooperation within the entire population over time using the mixed strategic approach, with the dilemma strength set at $r=0.05$. Initially, cooperation levels stand at 0.5 across all independent simulations, showing a downward trend thereafter. However, the degree of increase during the expansion phase and the stable cooperation levels vary significantly among simulations. To exemplify this variability, we highlight three representative episodes from the pool of 100 independent simulations.  In episode 1, cooperation only briefly enters the enduring period before facing extinction. In episode 2, cooperation experiences an expanding phase and eventually stabilizes at 26\%. Similarly, in episode 3, cooperation undergoes both the expanding and enduring periods, yet its enduring period is shorter, leading to stabilization at 55\%.

Figure \ref{fig6}  presents evolutionary dynamics snapshots under a mixed strategic approach for the three scenarios highlighted in Figure \ref{fig5}, with consistent parameter values. Initially, the strategies of all players are randomly distributed (left column of Figure \ref{fig6}). Given that defection (a low probability of choosing cooperation) is favored over cooperation (a high probability of choosing cooperation), the lattice quickly becomes dominated by a vast majority of players strongly inclined towards defection, with only a few cooperative players surviving in compact clusters. However, the probabilities of choosing cooperation within these cooperative clusters are uncertain. For instance, in episode 1, episode 2, and episode 3, we observe probabilities of 0.86, 0.92, and 0.99, respectively (second column of Figure \ref{fig6}). These distinctions may be related to the initial strategy distributions, as their initial fractions are the same but differ in their distributions. As the simulation progresses, we observe that while cooperative clusters form in all three episodes, those with low probabilities of choosing cooperation fail to resist the invasion of defectors and consequently go extinct soon after formation (first row of Figure \ref{fig6}). In contrast, clusters with intermediate probabilities of choosing cooperation are able to persuade neighboring players to adopt their strategy, thereby expanding their territories and ultimately leading to an intermediate level of cooperation (second row of Figure \ref{fig6}). Finally, clusters with high probabilities of choosing cooperation represent perfect cooperative clusters compared to those in the second row; they have a high ability to expand their territories, ultimately leading to the highest level of cooperation (third row of Figure \ref{fig6}). These dynamics are further illustrated in animated movies available at \url{https://osf.io/rfwcx/}, providing a dynamic visualization of how mixed strategic approach influences the evolution of cooperation.

\section{Conclusion and discussion}
In conclusion, our study examined the evolution of cooperation within the spatial prisoner's dilemma game framework, analyzing the influence of discrete, continuous, and mixed strategic approaches on cooperation dynamics separately. Our findings highlight that the mixed strategic approach introduces a degree of uncertainty into the evolution of cooperation, as evidenced by variations in stable cooperation levels across multiple independent simulations. This uncertainty is not only influenced by finite population size but also by the non-zero clustering coefficients of populations. These insights build upon previous research, which primarily focused on population size as a key factor influencing cooperation differences among strategic approaches ~\cite{zhong2012equilibrium}. 

The observed uncertainty in cooperation evolution is, to some degree, associated with the finite-size effect, as discussed in the literature~\cite{perc2017stability,perc2017statistical}. This effect underscores the potential for significant fluctuations within small systems, which can lead to the abrupt disappearance of actors, thereby yielding potentially misleading or unexpected outcomes. Given that many real-world systems are of finite size, our analysis highlights the substantial disparities in cooperation levels among the strategic approaches under consideration, particularly emphasizing the role of clustering coefficients in mitigating the uncertainty of cooperation evolution—a subtlety that becomes obscured in larger populations.

By definition, the discrete strategic approach involves deterministic actions, allowing players to choose between cooperation or defection with absolute certainty. In contrast, continuous and mixed strategic approaches relax this rigid assumption, permitting a probabilistic selection between cooperation and defection. Although the discrete approach facilitates simplified theoretical analysis, it potentially introduces biases in assessing the stability of cooperative systems. Shifting from deterministic actions, we observe increased uncertainty in the evolution of cooperation with continuous and mixed strategies. Notably, the mixed strategy incorporates stochastic elements into payoff calculations, determining immediate outcomes based on current actions, unlike the continuous strategy that relies on expected payoffs. This difference amplifies the uncertainty in cooperative evolution observed with mixed strategies compared to continuous ones. These observations collectively hint at a potential overestimation of the robustness of cooperative systems in real-world scenarios, underscoring significant uncertainty in cooperation evolution when employing more realistic mixed strategic approaches.

Our analysis specifically focuses on homogeneous networks, where each player uniformly interacts with the same number of opponents, thereby limiting our conclusions to such populations. This constraint is notable because real-world networks often exhibit a significant degree of heterogeneity~\cite{santos2005scale,santos2008social}, temporality~\cite{masuda2016guide,holme2012temporal}, higher-order structures~\cite{alvarez2021evolutionary,battiston2021physics}, among other characteristics~\cite{xu2023higher}. These characteristics suggest that extending our model to incorporate these aspects could yield insightful findings regarding the evolution of cooperation under more complex conditions. In particular, exploring the impact of scale-free networks with varying clustering coefficients~\cite{holme2002growing}, on the uncertainty of cooperation evolution presents a compelling avenue for future research.

\section*{Article information}

\paragraph*{Acknowledgements.} 
We acknowledge support from (i) a JSPS Postdoctoral Fellowship Program for Foreign Researchers (grant no. P21374), and an accompanying Grant-in-Aid for Scientific Research from JSPS KAKENHI (grant no. JP 22KF0303) to C.\,S., (ii) China Scholarship Council (no.~202308530309) and Yunnan Provincial Department of Education Science Research Fund Project (project no. 2023Y0619) to Z.H., and (iii) the grant-in-Aid for Scientific Research from JSPS, Japan, KAKENHI (grant No. JP 20H02314 and JP 23H03499) awarded to J.\,T.
\paragraph*{Author contributions.} 
C.\,S. and J.\,T. conceived research. Z.\,S. and Z.\,H. performed simulations. All co-authors discussed the results and wrote the manuscript.
\paragraph*{Conflict of interest.} Authors declare no conflict of interest.
\paragraph*{Data accessibility} The code used in the study to produce all results is freely available at  \url{https://osf.io/rfwcx/}.

\bibliographystyle{elsarticle-num}
\bibliography{biblio}

\end{document}